\documentclass[aps,prl,twocolumn,showpacs,groupedaddress]{revtex4}

\usepackage[latin1]{inputenc}
\usepackage{amsfonts}
\usepackage{amsmath}
\usepackage{graphicx,graphics}

\newcommand{\ex}[1]{\mathrm{e}^{#1}}

\newcommand{\ket}[1]{\left|{#1}\right\rangle}

\newcommand{\ea}{\emph{et al.}}

\begin{document}

\title{Entanglement Percolation in Quantum Complex Networks}
\author{Mart\'\i\ Cuquet}
\author{John Calsamiglia}
\affiliation{Grup de F\'{\i}sica Te\`orica, Universitat Aut\`onoma de
Barcelona, 08193 Bellaterra, Barcelona, Spain}
\date{\today}

\begin{abstract}
Quantum networks are essential to quantum information distributed
applications, and communicating over them is a key challenge. Complex networks
have rich and intriguing properties, which are as yet unexplored in the
quantum setting. Here, we study the effect of entanglement percolation as a
means to establish long-distance entanglement between arbitrary nodes of
quantum complex networks. We develop a theory to analytically study random
graphs with arbitrary degree distribution and give exact results for some
models. Our findings are in good agreement with numerical simulations and show
that the proposed quantum strategies enhance the percolation threshold
substantially. Simulations also show a clear enhancement in small-world and
other real-world networks.
\end{abstract}


\maketitle

A broad variety of natural and socioeconomic phenomena such as protein-protein
interactions, the brain, the  power-grid, friendship networks, the Internet
and foodweb, can all be modeled by graphs, i.e., by sets of nodes and edges
representing the relation between them. Complex networks (CN) cover the wide
range between regular lattices and completely random graphs. Their ubiquity
has triggered an intense research activity on CN involving applied works, but
also fundamental studies in theoretical physics and mathematics that aim at
unveiling their underlying principles. Understanding structural properties of
CN is very important as they crucially affect their functionality. For
instance, the topology of a social network affects the spread of information
or diseases, and the architecture of a computer network determines its
robustness under router failures. The appearance of a giant connected
component is a critical phenomenon related to percolation theory
\cite{grimmett_percolation_1989,dorogovtsev_critical_2008}. If, say, the
infection probability of a disease exceeds a critical value, an outburst of
the disease occurs infecting most of the population, whereas below the
critical value only a vanishing amount of nodes is affected. This threshold
and the size of the outburst are key properties that strongly depend on the CN
structure.

This Letter merges the field of CN with the also new and rapidly developing
field of quantum information (QI). Networks already have a prominent role in
QI, e.g., in the context of measurement-based quantum computation
\cite{raussendorf_one-way_2001}, in the characterization of graph states
\cite{hein_multiparty_2004}, and, more naturally, as the physical substrate of
nodes and channels for multipartite quantum communication protocols. Such a
``quantum internet'' \cite{kimble_quantum_2008} supports QI applications that
fill the technological gap between the already available bipartite
applications like quantum key distribution, and the appealing but still remote
quantum computer. The reason for studying quantum CN, as opposed to regular
lattices considered so far, is threefold. First, it is very plausible that
future quantum communication networks acquire a complex topology resembling
that of existing networks. This can certainly be the case if methods are
developed to use current communication networks at a quantum regime, or if new
quantum networks grow driven by similar sociopolitical mechanisms. Second, CN
have nontrivial topological features not found in regular lattices, e.g.,
robustness to random errors, that can be useful in the design of novel quantum
networks. Third, from a more theoretical perspective, although few qubit
states have been successfully characterized, an understanding of multipartite
entanglement of many-particle systems is still lacking. As we will show here,
some quantities which are exceedingly hard to calculate for states defined
over regular lattices can be carried with ease in CN (which are defined
through statistical properties).

The goal of entanglement percolation \cite{acin_entanglement_2007} is to
establish a maximally entangled state (which via quantum teleportation amounts
to a perfect single-use quantum channel) between two arbitrary nodes of a
network, where nodes are connected by partially entangled states. This can be
easily achieved between two neighbor nodes since a partially entangled state
can be converted into a maximally entangled state (for short \emph{singlet})
with a probability that depends on the initial amount of entanglement. In
1D networks, a singlet can only be established between two remote nodes with a
probability that decays exponentially with the distance between them
\cite{acin_entanglement_2007}. In contrast, higher dimensional networks have
an increasing number of possible connecting paths giving rise to a percolation
effect, i.e., to the appearance of a cluster spanning a significant fraction
$S$ of the network. Once a path of singlets is formed, a maximally entangled
state can be established between the end-nodes by means of entanglement
swapping. Thus, two arbitrary nodes can be connected by a singlet if both
belong to the same cluster. For large networks, this happens with a
probability $S^2$  independent of the distance between the nodes and of the
size of the network \cite{calsamiglia_spin_2005}, but that strongly depends on
its topology. Ac{\'\i}n \ea\ \cite{acin_entanglement_2007} found a type of
lattice where the value of the percolation threshold, i.e., the required
amount of initial entanglement, can be reduced by some quantum preprocessing:
\emph{local} quantum measurements are performed on some selected nodes of the
honeycomb lattice turning it into a triangular with a lower threshold. In
\cite{perseguers_entanglement_2008}, other local strategies were studied in
detail and entanglement percolation was shown to appear in few more 2D
lattices.

In this Letter, we study entanglement percolation in complex networks. We
first derive general results for random networks with arbitrary uncorrelated
degree distributions. From here, we can obtain exact values for the
percolation threshold and the probability $S$. We next complement these
results with numerical simulations on these and other paradigmatic CN, with
emphasis on those that mimic existing communication networks.

A network is naturally represented by a graph, which consists of $N$ vertices
and a number of edges that connect some pairs of vertices. The degree of a
vertex, $k$, is the number of edges connecting to it, and in general follows a
probability distribution $P(k)$. A connected component, or cluster, is a
subgraph where any two vertices are connected by at least one path and no more
vertices can be added without losing this property. If edges are occupied with
probability $p$, one can ask which is the distribution of connected component
sizes when this probability varies. For low values of $p$ many small clusters
exist. As $p$ increases clusters start to grow and join each other. In general
the cluster size is vanishingly small compared to the size of the network,
until a critical value of $p=p_{c}$ where a giant connected component (GCC)
starts to span it: in the asymptotic limit, this is the \emph{only} component
of finite relative size $S>0$ for all $p\geq p_{c}$.

We consider quantum networks, as in Ref.~\cite{acin_entanglement_2007}, whose
edges correspond to pairs of pure entangled states
$\left|\psi\right\rangle^{\otimes 2}$, where $\ket{\psi} = \sqrt{\lambda_0}
\ket{00} + \sqrt{\lambda_1}\ket{11}$ is a two qubit state with Schmidt
coefficients $\sqrt{\lambda_0}\ge\sqrt{\lambda_1}\ge 0$. Each partially
entangled state $\ket{\psi}$ can be converted into a singlet by LOCC with the
singlet conversion probability (SCP) $p = \min [ 1,
2(1-\lambda_0)]$~\cite{vidal_entanglement_1999}. If the two nodes share two
copies of $\ket{\psi}$, the probability that at least one of them is converted
into a singlet is of course $p_2 = 2p-p^2$. However,  they can do better since
the largest Schmidt coefficient of $\ket{\psi}^{\otimes 2}$ is $\lambda_0$,
and hence its optimal SCP is  $p_2 = \min [ 1, 2(1-\lambda_0^2) ] = \min ( 1,
2p-p^2/2 )$. From here on, ``with'' and ``without distillation'' respectively
mean that the optimal or the former sequential transformation is used.
\begin{figure}[b!]
  \includegraphics[width=\columnwidth]{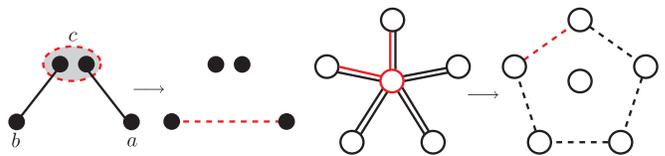}
  \caption{Entanglement swapping (left) and 5-swap transformation (right).
  Dots represent qubits, circles nodes, solid lines partially entangled states
  $\ket{\psi}$, and dashed lines correspond to states with the same SCP as
  $\ket{\psi}$.}
\label{fig:qswap}
\end{figure}
Applying the above singlet conversion to each edge of the network is called
classical entanglement percolation \cite{acin_entanglement_2007}. As already
mentioned, a new quantum feature appears when we allow for other local quantum
operations that transform the network geometry and thereby change the
percolation threshold. The basic ingredient is entanglement swapping (or
``swap'') illustrated in Fig.~\ref{fig:qswap}. The party at a central node $c$
performs a Bell measurement on two qubits, each of them belonging to states
$\ket{\psi}$ shared with different nodes ($a$ and $b$). After this operation,
the central qubits become disentangled from $a$ and $b$, but in return an
entangled state is established between $a$ and $b$, which on average has the
same SCP as $\ket{\psi}$. Note, however, that this operation cannot be
repeated using newborn edges since they are not in state $\ket{\psi}$. The
``$q$-swap'' operation that we use here performs the ``swap'' transformation
between successive pairs of neighbors of a central node of degree $q$,
effectively changing the initial $q$-star into a $q$-cycle
(Fig.~\ref{fig:qswap}). These $q$-swaps are made on nodes of certain degrees,
depending on the network structure with the only condition that no operation
is done on neighboring sites. Thus, the strategy uses only local information
of the network: the degree of the target node and the status of its neighbors.

We now proceed to present our results for general random graphs of arbitrary
(uncorrelated) degree distribution. We use the generating function formalism
\cite{callaway_network_2000, moore_exact_2000}. We begin by defining the
generating function of the degree distribution by $G_0(x) = \sum_{k=0}^\infty
P(k) x^k$. Since $P(k)$ is a probability, $G_0$ is well normalized,
$G_0(1)=1$, and convergent for $|x|\le1$. What makes generating functions
specially suited here is that it is straightforward to convolute probability
distributions; e.g., $G_{0}(x)^m$ generates the probability that $m$ arbitrary
nodes have a \emph{total} of  $k$  edges leading to them. The key probability
distribution in percolation is that of reaching a \emph{finite} component of
size $s$ when following a random edge to one of its ends, generated by
$H_1(x)$. The size is zero when the edge is not occupied. With probability
$p_2$ it is occupied, and the discovered vertex has degree distribution
$P_1(k) = kP(k)/\langle k \rangle$ generated by $\sum_{k=0}^\infty P_1(k)x^k =
x\frac{G_0^\prime(x)}{G_0^\prime(1)} \equiv xG_1(x)$. Thus, $s$ is one plus
the size of the new $k-1$ reachable components. The size of each of them is
again generated by $H_1(x)$, and their sum by $\sum_{k=0}^\infty P_1(k) [
H_1(x) ]^{k-1}$. Therefore, $H_1(x)$ fulfills the consistency equation
\begin{equation}
  H_1(x) = (1-p_2) + p_2 x G_1(H_1(x)).
  \label{eq:H1_original}
\end{equation}

It is crucial to notice that by restricting to finite $s$ we have explicitly
excluded the infinite GCC from $H_1(x)$ \footnote{We have assumed that
components are tree-like, since an edge exiting a finite component of size $s$
will connect back to the same component with a probability proportional to
$s/N\to 0$. This obviously does not hold for the GCC.}. The total probability
that an edge connects to a finite component is $u\equiv H_1(1)$, and due to
(\ref{eq:H1_original}) satisfies $u=1-p_2+p_2G_1(u)$. Below the critical
$p<p_{c}$ there is a unique solution $u=1$, while at $p\geq p_{c}$ a new
solution $u<1$ appears. Hence, the critical probability is the smallest value
of $p$ for which the second solution appears.

We can similarly compute the function $H_0(x)$ generating the probability that
a randomly chosen vertex belongs to a finite component of size $s$. Since $k$
edges emerge from this vertex with probability $P(k)$ we find $H_0(x) = x
G_0[H_1(x)]$. The probability that a randomly chosen vertex belongs to the GCC
is $S = 1- H_0(1)=1-G_0(u)$. From $H_0(x)$, we can compute many other relevant
properties, e.g., the mean component size below $p_{c}$,   $\langle s \rangle
= H_0^\prime (1)$, which behaves like a susceptibility diverging at the phase
transition.

Our goal now is to understand how quantum operations change the percolation
properties of the original network. Every particular $q$-swap can be
implemented (or not) with probability $\Pi_q$ (or $1-\Pi_q$) on nodes of
degree $q$. Giving the values for each $\Pi_{q}$ specifies the quantum
strategy. Now, instead of arriving to a vertex of degree $q$ connecting to
other $q-1$ components, after a $q$-swap operation we arrive to a cycle of $q$
nodes (including the one we are coming from) connected via links with SCP
equal to $p$ \footnote{$q$-swaps introduce cycles, and components are no
longer tree-like. However, such cycles do not overlap and can be considered
blocks of a tree-like component.}. The size of such cycle and its emerging
components is the sum of the probabilities that a connected string of $l$
vertices in the cycle is reached, multiplied by the generating function of the
emerging component sizes:
\begin{eqnarray}
  C_q(x)
  &=& \sum_{l=0}^{q-2} (l+1) p^l (1-p)^2 \left[ x G_1(\widetilde{H}_1(x)) \right]^l\nonumber
  \\
  &&+ \left[ qp^{q-1} (1-p) + p^q \right] \left[ x G_1(\widetilde{H}_1(x))
  \right]^{q-1}.
  \label{eq:Cq}
\end{eqnarray}
Therefore, the new $\widetilde{H}_1(x)$ is of the same form of
Eq.~(\ref{eq:H1_original}) plus a term $\widetilde{H}_{1,q}(x)$ for each
$q$-swap:
\begin{align}
  \widetilde{H}_1(x)
  &= 1 - p_2 + p_2 x G_1 ( \widetilde{H}_1(x) )
  + \sum_{q\ge2} \Pi_q \widetilde{H}_{1,q}(x)
  \label{eq:H1_modified} \\
  \widetilde{H}_{1,q}(x)
  &= 
  P_1(q)
  \left[
  (p_2-1)
  - p_2 x \left( \widetilde{H}_1(x) \right)^{q-1}
  + C_q(x)
  \right].\nonumber
  \label{eq:H1q_modified}
\end{align}
At this stage we can already calculate $\tilde{p}_c$ as the value of $p$ for
which there exists a solution $0<u=\widetilde{H}_1(1)<1$ to
\eqref{eq:H1_modified} at $x=1$. It is easy to convince oneself that each
separate contribution  $\widetilde{H}_{1,q}(1)$ either increases or lowers the
percolation threshold and therefore for the optimal strategy each $\Pi_q$ is
either 0 or 1.

In order to calculate the relative size of the giant component $\tilde S$ we
need to find the new $\widetilde{H}_0(x)$. Given a vertex $q$, there is a
probability $\eta_q$ that a $q$-swap can be performed on it, changing its
degree from $q$ to zero and hence
\begin{equation}
  \widetilde{H}_0(x)
  = x G_0 \left( \widetilde{H}_1(x) \right)
  + x \sum_{q\ge2} \Pi_q \eta_q P(q) \left[ 1 - \left(
  \widetilde{H}_1(x) \right)^q \right] .\nonumber
  \label{eq:H0_modified}
\end{equation}
From here, the order parameter $\tilde S = 1 - \widetilde{H}_0(1)$ can be
readily obtained by plugging in the solution of \mbox{$\widetilde{H}_1(1)=u$}.
The probability $\eta_q$ that no other $q^\prime$-swap was performed in its
neighbors can be computed up to any order in $p$ exactly \cite{cuquet2009}.
When only one type of $q$-swap is performed, a compact, approximate expression
exists $\eta_{q}\approx (1+\frac{q}{2}P_1(q))^{-1}$. In the following, we
study entanglement percolation enhancement in particular complex network
models.

The \emph{Bethe lattice} is an infinite regular tree where every vertex has
the same degree. Inserting $G_0(x)=x^q$ and $G_1(x)=x^{q-1}$ into
(\ref{eq:H1_original}) and (\ref{eq:H1_modified}) and solving the point at
which the solution $u<1$ appears gives $(q-1)^{-1}=p_2$ and $(q-1)^{-1} =
(1-p)^{-1} \{ 2p + p^q [ p(q-1)-(q+1) ] \}$ after $q$-swap is applied.
Therefore, $q$-swap gives always a better threshold except for an infinite 1D
chain ($q=2$).

\begin{figure}[b!]
  \includegraphics[width=\columnwidth]{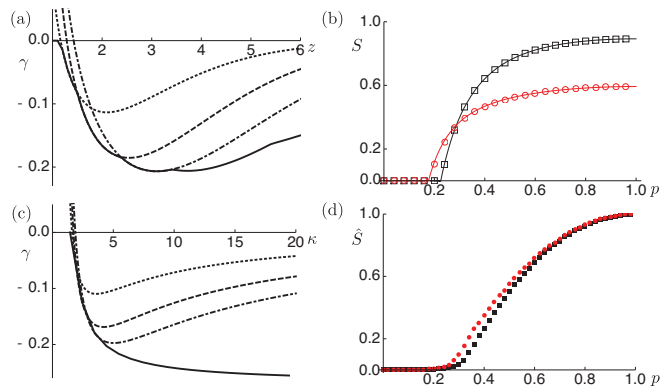}
  \caption{
  (a) Gain $\gamma$ in the ER model as a function of the mean degree $z$ after
  2-swap (dotted), 2,3-swap (dashed), 2,3,4-swap (dot-dashed) and optimal
  $q$-swaps (solid).
  (b) Size of the GCC as a function of $p$ in the $z=2.5$ ER network without
  distillation before (squares) and after 2,3-swap (circles). Points are
  simulation results for $N=10^6$.
  (c) Same as (a) for scale-free model with $\tau=1$ as a function of
  $\kappa$.
  (d) Same as (b) for \emph{WWW} \cite{albert_internet:_1999} with cutoff
  $k=15$ ($N\sim10^4$).
  }
  \label{fig:er_sf}
\end{figure}
\emph{Erd\H os--R\'enyi (ER) graphs} are maximally random graphs with the only
constraint $\langle k \rangle = z$. An ER network has $N$ vertices and each
pair of them holds an edge with probability $z/N$. The degrees follow a
Poisson distribution and $G_0(x)=G_1(x)=\exp [z(x-1)]$.  The original
threshold is given by $p_2=1/z$. After, e.g., the 2-swap and 3-swap
operations, the thresholds are respectively $\frac{1}{z} = p_2 + e^{-z} [ -p_2
+ z(2p-p^2) ]$ and $\frac{1}{z} = p_2 + ze^{-z} [ -p_2 + z(1+p-p^2) ]$. In
general the performance of different $q$-swaps depends on the mean degree $z$.
Figure~\ref{fig:er_sf}a shows the gain $\gamma = (\tilde{p}_c-p_c)/p_c$. The
threshold can improve by a 20\% (24\% without distillation), compared to the
3\% (16\% without distillation) of the honeycomb--triangular transformation in
\cite{acin_entanglement_2007}. The size of the GCC can be calculated exactly
from the above results \cite{cuquet2009} and is plotted in
Fig.~\ref{fig:er_sf}(b) showing perfect agreement with numerical simulations.
Note that $\tilde S$ attains smaller values for the transformed network even
at $p=1$ due to the target nodes that become detached after the $q$-swap.
Since one can always choose where to start applying the $q$-swaps, the
probability of connecting two arbitrary nodes is actually $\hat S^2$, where
$\hat S=\tilde S {S_{1}}/{\tilde{S}_{1}}$ is the probability that a node
belongs to the GCC excluding the detached nodes and $S_{1}$ is the GCC size at
$p=1$.

\emph{Scale-free networks}: Real world networks are not Poissonian but
typically have a power-law (scale-free) degree distribution. Such networks
have a GCC at all $p$, and a threshold appears only in scenarios where there
is a cutoff at high degrees due to, e.g., targeted attacks, physical
constraints or saturation effects. We have taken an exponential cutoff,
$P(k)=Ck^{-\tau}\ex{-k/\kappa}$, and find once again a significant improvement
in $p_{c}$, up to 25\% with distillation [Fig.~\ref{fig:er_sf}(c)]. We have
performed simulations for a real-world \emph{WWW} network
\cite{albert_internet:_1999}, where we have introduced a cutoff neglecting
nodes with $k\geq 15$ and, despite the finite-size effects, we clearly see a
decrease in  $p_{c}$ [Fig.~\ref{fig:er_sf}(d)].

\begin{figure}[t]
  \includegraphics[width=\columnwidth]{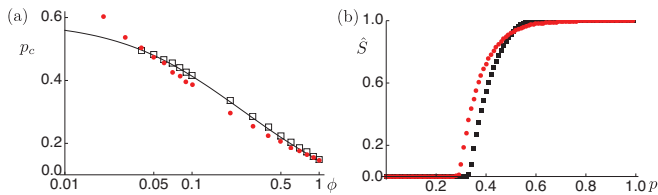}
  \caption{Numerical results for the SW model with 2-swap and distillation.
  (a) $p_c$ as a function of $\phi$ before (squares) and after (disks). Solid
  line is the classical analytic solution from \cite{moore_exact_2000}.
  (b) Size of the GCC as a function of $p$ before (squares) and after
  (circles) operations for $\phi=0.25$.
  }
  \label{fig:sw}
\end{figure}

A \emph{Small World} (SW) \cite{watts_collective_1998} is a network with
ordered local structure and high level of clustering but still with
surprisingly low average path length. We consider an SW model generated by
placing $N$ vertices in a 1D chain. Then $N$ additional random edges,
``shortcuts'', are added with probability $\phi$. In this case, degrees are
strongly correlated and thus the above analysis does not apply. Still,
simulations show that $q$-swaps can decrease $p_{c}$ (Fig.~\ref{fig:sw}).

To summarize, we have studied entanglement percolation in complex
networks---which are defined through their statistical properties and whose
exact structure is not necessarily known---and shown that by a strategy that
only uses local information we can decrease the percolation threshold
substantially (in some cases almost 1 order of magnitude more than found on
the honeycomb lattice). We have exactly solved models of random graphs with
arbitrary uncorrelated degree distribution. Moreover, we have numerically
explored the behavior in SW networks and the \emph{WWW}. In all cases, we find
that entanglement percolation, far from being restricted to particular
networks, is a quite general feature. Entanglement percolation is interesting
from the discipline of CN since quantum mechanics introduces a new paradigm
where networks can suffer non-trivial structural changes before the
percolation process starts. We believe our results open prospects for many
interesting synergies between the fields of quantum information and complex
networks.

An essential issue for entanglement percolation schemes is their stability
under noise. For amplitude damping, one can adapt the prescription in
\cite{broadfoot_entanglement_2009}. More general scenarios might require schemes
that, e.g., use multipartite entanglement, use global knowledge of the
network, concentrate different paths in a single one, or allow for the
generation of new (imperfect) bonds during the protocol. Finally, note that CN
can enjoy much better stability than lattices. In particular, in many CN, the
average path length scales with the size of the network as $\log N$, as
opposed to $N^{1/d}$ for $d$-dimensional lattices. Thus, the probability of
establishing entanglement between two arbitrary nodes has a linear falloff, as
opposed to a fatal exponential decay, and hence is non-negligible in large,
but finite, networks.

We are grateful to M. Bogu\~n\'a and A.V. Goltsev for discussions. We
acknowledge financial support from the Spanish MICINN, through the Ram\'on y
Cajal (J. C.) and AP2008-03048 (M. C.) and projects FIS2008-01236 and QOIT
(CONSOLIDER2006-00019), and the Generalitat de Catalunya Contract
No.~2009SGR985.

Note added.---Soon after the completion of this work, in
\cite{perseguers_quantum_2009} Perseguers \ea\ showed radically different
properties between quantum and classical Erd\H os--R\'enyi graphs, and also
stressed the interest in quantum CN. Interesting results for entanglement
distribution in regular lattices have also been recently reported in
\cite{perseguers_fidelity_2009}.


\begin{thebibliography}{10}

\bibitem{grimmett_percolation_1989}
G.~Grimmett,
\newblock {\em Percolation}.
\newblock {(Springer-Verlag, Berlin, 1989)}.

\bibitem{dorogovtsev_critical_2008}
S.~N. Dorogovtsev, A.~V. Goltsev, and J.~F.~F. Mendes,
\newblock {Rev.\ Mod.\ Phys.} {\bf 80}, 1275 (2008).

\bibitem{raussendorf_one-way_2001}
R.~Raussendorf and H.~J.~Briegel,
\newblock {Phys.\ Rev.\ Lett.} {\bf 86}, 5188 (2001).

\bibitem{hein_multiparty_2004}
M.~Hein, J.~Eisert, and H.~J.~Briegel,
\newblock {Phys.\ Rev.\ A} {\bf 69}, 062311 (2004).

\bibitem{kimble_quantum_2008}
H.~J.~Kimble,
\newblock{Nature (London)}, {\bf 453}, 1023 (2008).

\bibitem{acin_entanglement_2007}
A.~Acin, J.~I. Cirac, and M.~Lewenstein,
\newblock {Nature Phys.} {\bf 3}, 256  (2007).

\bibitem{calsamiglia_spin_2005}
J.~Calsamiglia \ea,
{Phys.\ Rev.\ Lett.} {\bf 95}, 180502 (2005);
L.~Hartmann \ea,
{J.\ Phys.\ B} {\bf 40}, S1 (2007);
K.~Kieling, T.~Rudolph, J.~Eisert,
Phys.\ Rev.\ Lett.\ {\bf 99}, 130501 (2007);
D.~E.\ Browne \ea,
{New J.\ Phys.} {\bf 10}, 023010 (2008).

\bibitem{perseguers_entanglement_2008}
S.~Perseguers \ea,
\newblock {Phys.\ Rev.\ A}  {\bf 77}, 022308 (2008);
J.~Lapeyre, J.~Wehr, and M.~Lewenstein,
\newblock {Phys.\ Rev.\ A} {\bf 79}, 042324 (2009).

\bibitem{vidal_entanglement_1999}
G.~Vidal.
\newblock {Phys.\ Rev.\ Lett.} {\bf 83}, 1046 (1999).

\bibitem{callaway_network_2000}
D.~S. Callaway \ea, 
\newblock {Phys.\ Rev.\ Lett.} {\bf 85}, 5468 (2000);
M.~A. Serrano and M.~Boguna,
\newblock {Phys.\ Rev.\ Lett.} {\bf 97}, 088701 (2006).

\bibitem{moore_exact_2000}
C. Moore, M. E. J. Newman
\newblock {Phys.\ Rev.\ E} {\bf 62}, 7059 (2000).

\bibitem{cuquet2009}
M.~Cuquet and J.~Calsamiglia.
\newblock{(in preparation)}.

\bibitem{albert_internet:_1999}
R.~Albert, H.~Jeong, and A.~Barab{\'a}si.
\newblock {Nature (London)} {\bf 401}, 130 (1999).

\bibitem{watts_collective_1998}
D.~J. Watts and S.~H. Strogatz,
\newblock {Nature (London)} {\bf 393}, 440 (1998).

\bibitem{broadfoot_entanglement_2009}
S.~Broadfoot, U.~Dorner, and D.~Jaksch,
\newblock {arXiv:0906.16221v2}.

\bibitem{perseguers_quantum_2009}
S.~Perseguers \ea,
\newblock{arXiv:0907.3283v1}.

\bibitem{perseguers_fidelity_2009}
S.~Perseguers,
\newblock{arXiv:0910.1459};
S.~Perseguers \ea,
\newblock{arXiv:0910.2438}.
\end{thebibliography}
\end{document}